# Machine Learning Surrogates for Predicting Response of an Aero-Structural-Sloshing System

S. Srivastava*, M. Damodaran** and B.C. Khoo#

*National University of Singapore*
*Faculty of Engineering*
*9 Engineering Drive 1, Singapore 117575*

## ABSTRACT

This study demonstrates the feasibility of developing machine learning (ML) surrogates based on Recurrent Neural Networks (RNN) for predicting the unsteady aeroelastic response of transonic pitching and plunging wing-fuel tank sloshing system by considering an approximate simplified model of an airfoil in transonic flow and sloshing loads from a partially filled fuel tank rigidly embedded inside the airfoil and undergoing a free unsteady motion. The ML surrogates are then used to predict the aeroelastic response of the coupled system. The external aerodynamic loads on the airfoil and the two-phase sloshing loads data for training the RNN are generated using open-source computational fluid dynamics (CFD) codes. The aerodynamic force and moment coefficients are predicted from the surrogate model based on its motion history. Similarly, the lateral and vertical forces and moments from fuel sloshing in the fuel tank are predicted using the surrogate model resulting from the motion of the embedded fuel tank. Comparing the free motion of the airfoil without sloshing tank to the free motion of the airfoil with a partially-filled fuel tank shows that the effects of sloshing on the aeroelastic motion of the aero-structural system. The effectiveness of the predictions from RNN are then assessed by comparing with the results from the high-fidelity coupled aero-structural-fuel tank sloshing simulations. It is demonstrated that the surrogate models can accurately and economically predict the coupled aero-structural motion.

## Nomenclature

| | | | |
|---|---|---|---|
| $m$ | mass of wing section ($kg/m$) | $m_{tot}$ | combined mass of wing section and enclosed fluid ($kg/m$) |
| $S_\alpha$ | static imbalance of wing section ($kg/m$) | $S_{\alpha,tot}$ | static imbalance of wing section with enclosed fluid ($kg/m$) |
| $M_\infty$ | freestream Mach number | $V_f$ | speed index |
| $C_L$, $C_M$ | aerodynamic lift and moment coefficient | $F_X$, $F_Y$, $M_Z$ | horizontal force ($N$), vertical force ($N$) and moment ($N$-$m$) due to sloshing |
| $\omega_\alpha$, $\omega_h$ | Pitching and plunging frequencies | $mse$ | mean squared error |

___________________

*Graduate Student, Dept. of Mechanical Engineering; shashank.srivastava@u.nus.edu

** Senior Research Scientist, Temasek Laboratories; Associate Fellow AIAA; tslmura@nus.edu.sg

# Professor of Mechanical Engineering and Director Temasek Laboratories; tslhead@nus.edu.sg

## I. Introduction

The advent of machine learning (ML) approaches in science and engineering has been gaining a rapid pace in recent times and a recent workshop on Scientific ML by Baker et al. [1] outlined recommendations for enhancing its predictive power by augmenting it with scientific data. In the context of computational engineering predictions, this primarily implies developing efficient, economical, fast and reliable ML prediction tools while not compromising on the richness of the high-fidelity continuum-physics models. It is in this context that the present study explores the feasibility of developing machine learning surrogates for predicting unsteady aerodynamic characteristics of aero-structural systems in which the understanding of fluid-structure interaction is essential. The most common approach to model these systems is to use high fidelity continuum-physics based models embedded in various flow and structural solvers which can be computationally expensive if the system is complex and requires multi-physics interaction to realize the final objectives. A cheaper alternative will be to either use low-fidelity approaches based on gross simplifications or model order reduction approaches built using data from high-fidelity models and more recently machine learning based predictive approaches as in Swischuk et al. [2] and Dupuis et al. [3]. The problem of interest to motivate this study is the standard unsteady motion of an airfoil in a combined pitching and plunging motion immersed in an external transonic flow except that in this study a rectangular fuel tank embedded inside the airfoil and which is half-filled with fuel which will slosh around resulting in sloshing loads during the motion. The effects of fuel sloshing in aero-structural systems in transonic flights have been addressed in Cazier et al. [4] and Chiu and Farhat [5]. Despite these studies, there is, in general, a paucity of work in assessing the effects of sloshing on unsteady motion of the airfoil. This can be attributed to the computational expense required to run coupled aeroelastic-multiphase sloshing solvers, as well as the general lack of multi-physics solvers to tackle such problems.

Sloshing of liquids inside partially filled containers plays an important role in determining the dynamics of enclosing vessel such as marine LNG carrier, ground vehicles, and aircrafts with partially fuel tanks are prone to the effects of sloshing. Abrahamson [6] and Luskin [7] coupled sloshing fuel tank to space vehicle and aircrafts respectively using simplified sloshing models to represent the dynamic behaviour of a complicated phenomenon. Low order models for external aerodynamics such as Theodorsen's method [8] fail to capture the nonlinearities in the flowfield and are valid for low speed, low-angle of attack, linear flows. Similarly, low fidelity sloshing models such as equivalent mechanical systems outlined in Ibrahim [9] are unable to capture flow field nonlinearities, which are well captured by high-fidelity CFD analysis. Accurate flow solutions can be obtained by coupling CFD based aeroelastic solver with a multiphase flow solver in time-domain but the prohibitive computational effort and costs render it as an undesirable candidate for iterative design purposes. This motivates the development of low-cost surrogate model which can efficiently and accurately predict the dominant dynamics of coupled aeroelastic-sloshing multiphysics system which incorporate nonlinear flow field effects.

One route for this is via model order reduction as in Lucia et al. [10] and Dowell and Hall [11] which provide a comprehensive overview of several reduced-order techniques such as Harmonic Balance, Volterra theory and Proper Orthogonal Decomposition (POD) for aeroelastic applications. These approaches use linear system identification concepts to obtain a reduced-order model and hence such methods based on state-space approach cannot accurately capture non-linearities of the sloshing model, the large amplitude vibrations and limit cycle oscillations, which require specialized methods for non-linear system identification. The other route which is gaining momentum in recent years is via machine learning. Faller and Schreck [12] proposed a recurrent Multi-Layer-Perceptron Neural Network (MLP-NN) to predict unsteady loads for aeroelastic applications which was followed up by Voitcu and Wong [13] and Mannarino and Mantegazza [14] leading to a systems model for describing the aeroelastic behavior of airfoils and wings based on non-recurrent MLP-NN. The dynamic loads of fuel sloshing in a fuel tank not only depends on its current state, but also on the previous states and inputs since the fluid is always in a transition. In order to include dynamic memory effects temporal derivatives of the excitation signal are to be added to the input vector of the neural network. Neural networks have been shown to offer a powerful tool in modeling nonlinear systems over a compact set rather than a small neighborhood around the dynamically linearized steady state approach for linear systems. It provides a powerful tool for learning complex input-output mappings and has simulated many studies for identification of dynamic systems with unknown nonlinearities as outlined in a number of studies such as Narendra and Parthsarathy [15] and Elanayar and Shin [16]. In this study, machine learning surrogates based on the Recurrent Neural Network (RNN) outlined in Mandic [17] are developed to predict both external aerodynamic loads and sloshing loads for describing the coupled motion of the combined airfoil and fuel tank system. The methodology is demonstrated on a 2-DOF system which approximates a wing-section enclosing a partially filled fuel-tank. Training data for the RNN, rich in dominant dynamic response of the underlying physical system, generated using high fidelity CFD analysis. These machine learning surrogates are explicitly coupled in time domain to replicate the unsteady motion of the aero-structural system under consideration in this study. Section II outlines the aero-structural model for capturing the unsteady response in external transonic flow using high-fidelity compressible flow solvers. Section III outlines the flow model for fuel sloshing in a rectangular fuel tank embedded inside the aero-structural model. Section IV briefly outlines the coupling of these high-fidelity models to compute the actual response (i.e. the ground truth) and to develop high-fidelity physics-based data for training the RNN to form the machine learning surrogate for predicting the unsteady response of the aero-structural system. Section V discusses the results and reliability of the predictions of the RNN surrogate model in relation to the ground truth predicted by the high-fidelity physics-based model.

## II. Aero-Structural Model

The aero-structural model for an airfoil undergoing a pitching and plunging motion in an external transonic flow as shown in Fig 1. is modelled as follows:

$$[M]\{\ddot{q}\}+[K]\{q\}=\{F_{aero}\} \quad (1)$$

where

$[M] = \begin{bmatrix} m & S_\alpha \\ S_\alpha & I_\alpha \end{bmatrix}$, $[K] = \begin{bmatrix} k_h & 0 \\ 0 & k_\alpha \end{bmatrix}$, $\{q\} = \begin{Bmatrix} \bar{h} \\ \alpha \end{Bmatrix}$, $[F_{aero}] = \begin{bmatrix} -L \\ M_{ea} \end{bmatrix}$, $\bar{h} = \dfrac{h}{b}$, where $m$ is the mass of the

wing section per unit span of the wing (kg/m), $S_\alpha = x_a mb$ is the static imbalance (kg/m), $b$ is the semi-chord length of the wing section (m) and $x_\alpha$ is the distance of the center of gravity (cg) behind the mid-chord point of the wing section (m), $I_\alpha$ is the polar moment of inertia about the elastic axis which can also be written as $I_\alpha = mbr_\alpha^2$ where $r_\alpha$ is the radius of gyration of the wing section about the elastic axis, $k_h$ is the stiffness coefficient associated with plunging (N), $k_\alpha$ is the torsional stiffness associated with pitching (N), $h$ is the plunge displacement of the elastic axis (labelled as point *ea* in Fig.1) measured relative to the origin of the coordinate system and positive along the negative *y*-direction (m), $\alpha$ is the pitch displacement measured relative to the initial angle of attack about which the unsteady pitching is initiated (rad), and $L$ and $M_{ea}$ are respectively the vertical force and pitching moment about the elastic axis. These non-dimensional form of Eqn. (1a) is as follows:

$$[\bar{M}]\{\ddot{q}\} + [\bar{K}]\{q\} = \{\bar{F}_{aero}\} \tag{1b}$$

where $[\bar{M}] = \begin{bmatrix} 1 & x_\alpha \\ x_\alpha & r_\alpha^2 \end{bmatrix}$ and $[\bar{K}] = \begin{bmatrix} \left(\frac{\omega_h}{\omega_\alpha}\right)^2 & 0 \\ 0 & r_\alpha^2 \end{bmatrix}$, $\omega_\alpha = \sqrt{\frac{k_\alpha}{I_\alpha}}$ and $\omega_h = \sqrt{\frac{k_h}{m}}$ are the frequencies associated with the pitching and plunging respectively and $V_f = U_\infty (b\omega_\alpha)^{-1}$ is the speed index, $\{\bar{F}_{aero}\} = \frac{\rho_\infty U_\infty^2}{m\omega_\alpha^2}\begin{Bmatrix} -c_L \\ 2c_M \end{Bmatrix} = \frac{4}{\pi \mu k_c^2}\begin{Bmatrix} -C_l \\ 2C_m \end{Bmatrix}$ is the non-dimensional aerodynamic load, $k_c = U_\infty^{-1}\omega_\alpha c$ is the reduced frequency, $\mu = \frac{m}{\pi \rho_\infty b}$ is the mass ratio and $\{q\} = \begin{Bmatrix} h/b \\ \alpha \end{Bmatrix}$ are displacement vectors for plunge and pitch.

The airfoil considered for this study is the NACA 0012 airfoil which is subjected to a forced harmonic pitching and plunging motion initially for a few cycles before it is set to free pitching and plunging. The pitch and plunge are measured from the inertial frame fixed at the airfoil leading edge at *t=0* in the direction shown in Fig. 1.

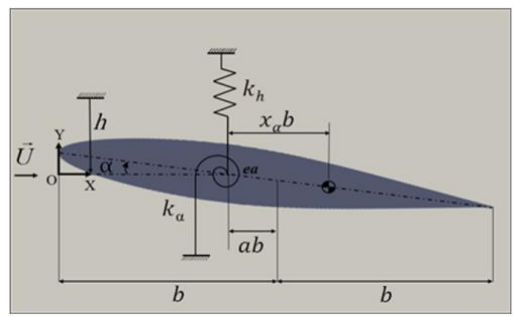

**Figure 1: Airfoil in Pitch and Plunge Motion**

The unsteady external transonic flow field is modeled using Euler equation cast into ALE formulation

$$\frac{\partial}{\partial t}\int_{\Omega(t)} \vec{U}\, d\Omega + \int_{\partial \Omega(t)} \vec{F}\vec{n}\, dS = 0 \tag{1c}$$

where control volume $\Omega(t)$ and its boundary $\partial\Omega(t)$ are time-dependent, $\vec{U}$ denotes the flow variables, $\vec{F}^c(U)$ is convective flux term and $\vec{\dot{u}}_m$ is the mesh velocity vector. These vectors are defined as

$$\vec{U} = \begin{bmatrix} \rho \\ \rho\vec{V} \\ \rho E \end{bmatrix} \text{ and } \vec{F} = \begin{bmatrix} \rho(\vec{V} - \vec{\dot{u}}_m)\vec{n} \\ \rho\vec{V}(\vec{V} - \vec{\dot{u}}_m)\vec{n} + p\vec{n} \\ \rho E(\vec{V} - \vec{\dot{u}}_m)\vec{n} + p\vec{V}\vec{n} \end{bmatrix} \quad (1d)$$

where $\rho, \vec{V}, E, p$ are the density, the velocity vector, fluid enthalpy and static pressure respectively. The sign convention of the aerodynamic load coefficients is set to be consistent with that used in the open source compressible flow solver *SU2* as outlined in Economon et al. [18] which is used to compute the external transonic aerodynamic flow fields for the aero-structural system. This model corresponds to the aero-structural motion of the airfoil with no fuel tank with sloshing fuel and is used to generate training samples for the predictive neural network model and for computing the external transonic aerodynamic flow fields.

The structural parameters for airfoil model used are are $x_\alpha = 1.8$, $r_\alpha^2 = 3.48$, $\omega_h = 100$ *rad/s*, $\omega_\alpha = 100$ *rad/s* and $\mu = 60$. From the flow model one could estimate the time variation of the aerodynamic force and moment coefficients.

## III. Fuel-Sloshing Model

Fuel sloshing in a partially filled rectangular fuel tank is modelled using a multi-phase fluid model based on the incompressible Navier-Stokes Equations consisting of the continuity and momentum equations as follows:

$$\int_{\Omega(t)} (\vec{U} \cdot \vec{n}) d\Omega = 0 \quad (2a)$$

$$\frac{\partial}{\partial t}\int_{\Omega(t)} \vec{U} d\Omega + \int_{\partial\Omega(t)} (\vec{U}(\vec{U} \cdot \vec{n})) dS = -\frac{1}{\rho}\int_{\partial\Omega(t)} (\nabla p - \mu(\nabla . \nabla)\vec{U}) dS + \int_{\Omega(t)} (\vec{f}_B + \vec{f}_v) d\Omega \quad (2b)$$

where $\vec{U}$ denotes the velocity of fluid relative to the tank, $\Omega(t)$ and $\partial\Omega(t)$ denotes are time-dependent control volume and its boundary, $p$ the pressure, $\rho$ and $\mu$ the density and viscosity, respectively, $\vec{f}_B$ represents the external body forces per unit mass for the fluid due to gravity and $\vec{f}_v$ represents body force per unit mass on the fluid influenced by tank motion. Fuel sloshing in the tank consists of two phases namely fuel and the vapor regions in the tank. The volume of fluid (VOF) method of Hirt and Nicholas [19] is used for tracking the volume of these phases in a fixed Eulerian mesh for the internal flow in the tank. In this method, a single set of momentum equations is shared by the fluid phases and the volume fraction of each fluid is tracked throughout the domain. A scalar function *f* is used to characterize the free surface deformation, the value of which is set based upon the fluid volume fraction of a cell. The volume fractions are updated by the equation written as

$$\frac{D}{Dt}\int_{\Omega(t)} f d\Omega = \int_{\Omega(t)} \frac{\partial f}{\partial t} d\Omega + \int_{\partial\Omega(t)} (f(\vec{U} \cdot \vec{n})) dS = 0 \quad (3)$$

After computation of the volume fractions in each cell, the equivalent density $\rho_{cell}$ and viscosity $\mu_{cell}$ are estimated as $\rho_{cell} = \rho_{gas} + f(\rho_{fuel} - \rho_{gas})$ and $\mu_{cell} = \mu_{gas} + f(\mu_{fuel} - \mu_{gas})$ where $\rho_{gas}$ and $\mu_{gas}$ are the density and viscosity of the vapor respectively, while $\rho_{fuel}$ and $\mu_{fuel}$ are the density and viscosity of the fuel in the tank. The fuel sloshing problem is solved in time-domain and at each time-step, the sloshing forces and moments acting on the tank wall is obtained by integrating the pressure fields and shear forces along the tank walls. The sloshing of the fuel in a partially filled tank is computed using the *InterDyMFOAM* library which implements this VOF method in *OpenFOAM* [20]. The pressure and shear forces from the computation are integrated over the wetted area of the fuel tank wall as shown in Fig. 2 and are used to estimate the sloshing loads as follows:

$$F_{x,sl} = \int_{-h/2}^{h/2} p_E(y) w \, dy - \int_{-h/2}^{h/2} p_W(y) w \, dy + \int_{-l/2}^{l/2} \tau_S(x) w \, dx + \int_{-l/2}^{l/2} \tau_N(x) w \, dx \qquad (4a)$$

$$F_{y,sl} = \int_{-l/2}^{l/2} p_N(x) w \, dx - \int_{-l/2}^{l/2} p_S(x) w \, dx + \int_{-h/2}^{h/2} \tau_W(y) w \, dy + \int_{-h/2}^{h/2} \tau_E(y) w \, dy \qquad (4b)$$

The corresponding moments about geometric centre of the tank are estimated as follows:

$$\begin{aligned} M_z = &\int_{-h/2}^{h/2} p_E(y) w (y - y_{EA}) dy + \int_{-l/2}^{l/2} p_N(x) w (x - x_{EA}) dx - \int_{-l/2}^{l/2} p_W(y) w (y - y_{EA}) dy \\ &- \int_{-l/2}^{l/2} p_S(x) w (x - x_{EA}) dx + \int_{-h/2}^{h/2} \tau_W(y) w (x - x_{EA}) dy + \int_{-l/2}^{l/2} \tau_N(x) w (y - y_{EA}) dx \\ &- \int_{-h/2}^{h/2} \tau_E(y) w (x - x_{EA}) dy - \int_{-l/2}^{l/2} \tau_S(x) w (y - y_{EA}) dx \end{aligned} \qquad (4c)$$

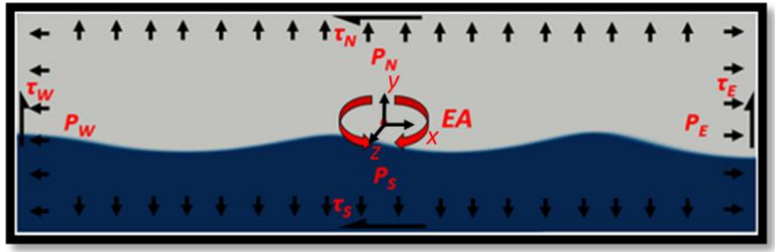

**Figure 2: Fuel tank with sloshing fluid**

The sloshing forces are the vector sum of the forces on the wall, and hence there is no need to take special care of inertial forces. The local axis, *xyz*, is attached to the geometric centre of the fuel tank and the sloshing loads namely, $F_x$, $F_y$ and $M_z$, are computed relative to this frame of reference. The sloshing loads are then projected in the inertial frame i.e. *XYZ* to form the coupled aero-structural model of Fig 1 using the transformation,

$$F_{X,sl} = F_{x,sl} \cos\alpha + F_{y,sl} \sin\alpha \qquad (5a)$$

$$F_{Y,sl} = F_{y,sl} \cos\alpha - F_{x,sl} \sin\alpha \qquad (5b)$$

$$M_{OZ,sl} = M_{oz,sl} \qquad (5c)$$

## IV. Coupled Aero-Structural and Fuel Tank Sloshing Model

The coupled aero-structural system consists of the pitch-plunge airfoil and a partially filled fuel tank rigidly attached to and embedded inside the airfoil as shown in Fig 3. The fuel tank is attached to the airfoil such that the geometric centre of the tank coincides with the elastic axis of the airfoil.

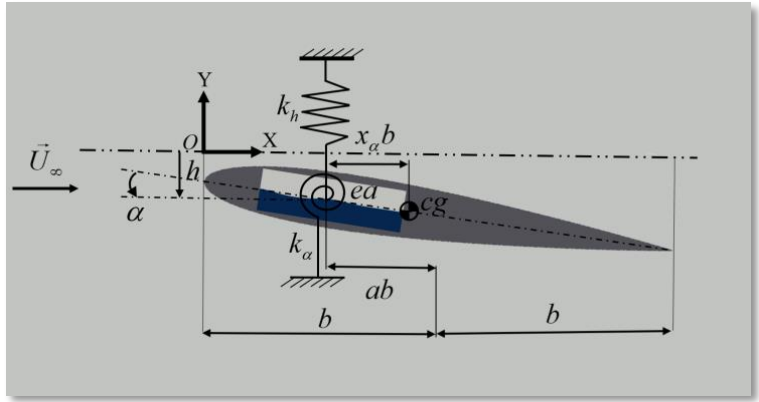

**Figure 3. Pitch-plunge airfoil with rigidly attached embedded partially filled fuel tank**

As a result of embedding the fuel tank in the airfoil, Eqn. (1) is modified to accommodate the additional fuel mass and the dynamic sloshing forces and moments as follows:

$$[M]\{\ddot{q}\} + [K]\{q\} = \{F_{aero}\} + \{F_{slosh}\} \tag{6}$$

where

$$[M] = \begin{bmatrix} m_{tot} & S_{\alpha,tot} \\ S_{\alpha,tot} & I_\alpha \end{bmatrix}, [K] = \begin{bmatrix} k_h & 0 \\ 0 & k_\alpha \end{bmatrix}, \{q\} = \begin{Bmatrix} \bar{h} \\ \alpha \end{Bmatrix}, \{F_{aero}\} = \begin{Bmatrix} -L \\ M_{ea} \end{Bmatrix}, \{F_{slosh}\} = \begin{Bmatrix} F_{Y,ea} \\ M_{Z,ea} \end{Bmatrix},$$

$m_{tot} = m + m_f$ is the combined mass of the airfoil $m$ and the fuel mass $m_f$ in the fuel tank (kg/m), $S_{\alpha,tot}$ is the static imbalance due to combined mass (kg) and all other parameters are defined same as in uncoupled aeroelastic formulations. It is assumed that shift in centre of gravity of the sloshing fluid is very small and hence is not incorporated in the equation of motion. The sloshing forces and moments $F_{Y,ea}$ and $M_{Z,ea}$ in $\{F_{slosh}\}$ are computed at the elastic axis at each structural time step defined by $\tau_{stuct} = \omega_\alpha t$. The non-dimensional form of Eqn. (6) consistent with Eqn. (1b) is as follows:

$$\frac{m_{tot}}{m}\begin{bmatrix} 1 & x_\alpha \\ x_\alpha & r_\alpha^2 \end{bmatrix}\begin{Bmatrix} \ddot{\bar{h}} \\ \ddot{\alpha} \end{Bmatrix} + \begin{bmatrix} \left(\frac{\omega_h}{\omega_\alpha}\right)^2 & 0 \\ 0 & r_\alpha^2 \end{bmatrix}\begin{Bmatrix} \bar{h} \\ \alpha \end{Bmatrix} = \frac{V^{*2}}{\pi}\begin{Bmatrix} -C_L \\ 2C_M \end{Bmatrix} + \frac{1}{mb\omega_\alpha^2}\begin{Bmatrix} F_{Y,ea} \\ 2M_{Z,ea} \end{Bmatrix} \tag{7}$$

Equation (7) is solved by the Rayleigh-Ritz modal approach for solving the generalized eigenvalue problem as outlined in Alonso and Jameson [21]. The structural displacement vector is decomposed first, followed by pre-multiplication by $[\phi]^T$, yielding a set of equations in generalized coordinates as follows:

$$\{q\} = [\phi]\{\eta\} \tag{8a}$$

$$[\phi]^T[M][\phi]\{\ddot{\eta}\}+[\phi]^T[K][\phi]\{\eta\}=[\phi]^T(\{F_{aero}\}+\{F_{slosh}\}) \qquad (8b)$$

$$\ddot{\eta}_i+\omega_i^2\eta_i=Q_i \qquad (8c)$$

where $\{\phi\}_i^T[K]\{\phi\}_i=\omega_i^2$ and $Q_i=\{\phi\}_i^T(\{F_{aero}\}_i+\{F_{slosh}\}_i)$ and $i=1,2$ corresponds to the plunging and pitching modes which are solved by splitting this second-order ODE into two first-order ODEs as follows:

$$\{\dot{X}_i\}=[A]\{X_i\}+\{F_i\} \qquad (9)$$

where, $X_i=\begin{Bmatrix}x_i\\\dot{x}_i\end{Bmatrix}, A=\begin{bmatrix}0 & 1\\-\omega_i^2 & 0\end{bmatrix}$ and $\{F_i\}=\begin{Bmatrix}0\\Q_i\end{Bmatrix}$. Equation (9) is integrated in time for the pitch and plunge displacements at the current time step and the global displacements of the airfoil are used to estimate the new displacements ($h$ and $\alpha$). The sloshing model is then solved for this new position and subsequent sloshing loads are computed. In this manner, the coupled equations for the aeroelastic surrogate and sloshing surrogate model marches on in time and the unsteady response of the aero-structural system evolves with time. The external aerodynamic forces on the airfoil are computed using *SU2* [18] and the sloshing forces in the fuel tank inside the embedded fuel tank are computed using the *InterDyMFOAM* library in *OpenFOAM* [20]. The execution of these different solvers in a time-accurate manner requires the coupling of these independent solvers with the structural model for facilitating data transfer via an interface software during the computation. The aeroelastic solver and the incompressible two-phase solver for fuel sloshing in the tank interact via the transfer of forces and moments and the corresponding structural displacements in the time domain computed at each structural time step. These solvers are coupled using an open source *preCICE* (Partitioned Multi-Physics Simulations) library outlined in Ref. [22] as shown in the simulation workflow diagram in Fig 4. The existing *preCICE* adapters for *SU2* and *OpenFOAM* have been modified in Srivastava et al. [23] to facilitate this coupling.

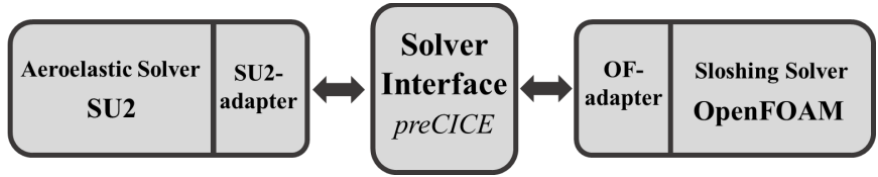

**Figure 4: Workflow in *preCICE* based modified SU2 and OpenFOAM adapters**

The structural time step is governed by the aeroelastic solver i.e. *SU2* and the pitch-plunge data and sloshing forces and moments are exchanged using the *preCICE Solver Interface* at each structural time step. The existing *SU2 adapter* facilitates writing of the fluid forces and moments to the *preCICE* Solver Interface which can read the structural displacements is modified to read integrated sloshing forces and moments and write structural displacements in form of pitch and plunge motion on to the *preCICE* Solver Interface. The *OpenFOAM adapter* is modular in nature and *preCICE* adapter is non-intrusive in nature. This renders higher flexibility for the modification of *preCICE* adapter for the current problem. The *interDyMFoam* solver used for fuel sloshing computation in internally attached fuel tank is accordingly modified in the adapter. The *preCICE* Solver Interface writes the displacement values to *OpenFOAM adapter* and reads the fluid forces and moments from it.

## V. Machine Learning Based Surrogates for Predicting Unsteady Responses

The construction of a machine learning surrogate for a reliable economic prediction of the unsteady response of the aero-structural system shown in Fig. 3 and outlined in Sections 1-IV, requires special care. Certain knowledge of system dynamics and response is necessary to construct and test a neural network architecture which can efficiently predict its future response. The traditional neural network assumes a black-box approach and generally utilizes a feedforward structure for input-to-output mapping.

The unsteady motion of the airfoil and the internal fuel sloshing is influenced not only by the current state of the system, but also by its previous states. Hence this requires the incorporation of memory in the ML surrogate model for storing historical data which is a major shortcoming in many traditional neural networks. This shortcoming can be overcome by deploying a *Recurrent Neural Network* (RNN) which not only allows nonlinear modelling of training data, but also processes temporal data for predictive modelling. RNNs are designed natively to allow deep learning networks to process sequences of data as shown in Fig 5. The state, $s$ is an indicator of the temporal history i.e. it represents the effects of all the previous inputs and predicted outputs on the current time step. A single time step of the input is supplied to the network and the current state is calculated by a nonlinear transformation, $g$ which is a function of the present input and the previous state i.e.,

$$S_t = g(x_t, S_{t-1}) \tag{10a}$$

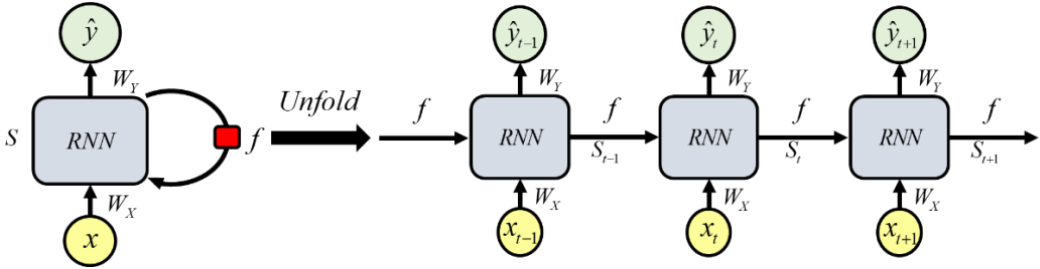

**Figure 5: Recurrent neural network cells in abstract and unfolded representation**

The current state, $S_t$ becomes the previous state for the next step and the process is repeated until the desired number of steps are reached. The final current state is then used to calculate the network output, $\hat{y}_t$ as follows

$$\hat{y}_t = f(x_t, S_t) = f(x_t, x_{t-1}, S_{t-1}) \tag{10b}$$

where $f$ is a nonlinear function that encodes the underlying dynamics of the system learned from the data and can make further predictions. The activation function used in the hidden neurons for this study is *rectified linear unit* (ReLU), defined by $f(x) = \max(0, x)$ and outlined in Ramachandran et al. [24]. The linear unit overcomes the problem of vanishing gradient [17] because of its partial linear nature and hence it is a good candidate for training networks with more than one hidden layer. It also performs a nonlinear operation on the input and state vectors, and its combined output is reflected in the encoding function $f$. The inputs and outputs of the sequential data through the hidden units in an RNN architecture at different time steps are considered as if they are discreet inputs and outputs of different neurons with the same weights and biases. This is evident from Fig 5 where the same weight, $W_x$, is used for all the sequential outputs of the representative RNN cell facilitating sequential learning. A detailed representation of the densely connected multi-layer (deep) recurrent neural network in the context of the problem of interest is shown in Fig 6(a) for the overall aero-structural system and in Fig 6(b) for the sloshing loads in the fuel tank.

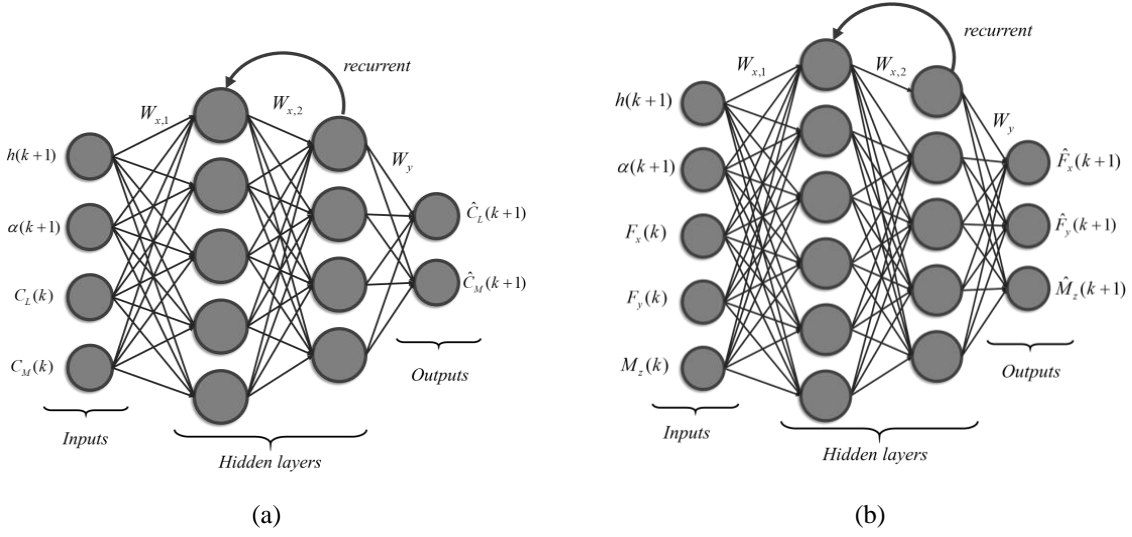

**Figure 6: Densely connected deep recurrent neural network in multiple-input multiple-output configuration for (a) the overall aeroelastic model, and (b) sloshing loads in the fuel tank model**

The computational nodes are connected through weights that are collectively represented by $W_{x,1}$ and $W_{x,2}$ for the first and second hidden layers respectively in Fig 6. The weights are initialized with random values and then adapted during the training process to improve network performance. For the aeroelastic prediction, the inputs are given in form of temporal sequences of airfoil plunge and pitch displacements i.e. $h$ and $\alpha$ and also the network outputs i.e. lift coefficient, $C_L$ and pitching moment coefficient $C_M$ at the previous instant. The sequential input data at the present time step, represented by $h(k+1)$ and $\alpha(k+1)$ and load coefficients sequences offset by one time step represented by $C_L(k)$ and $C_M(k)$ in Fig 6 (a) and the same notation is followed for sloshing loads shown in Fig 6 (b). These inputs are passed through the activation functions in the neurons in each hidden layer and the outputs obtained are the lift and moment coefficients at the present time step. It is important to note that for all temporal inputs in a batch of training data, the weights of the neurons remain the same.

The network is *trained* by minimizing *loss functions* which are essentially the mean squared error (*mse*) of the difference between the output predicted by the network $\hat{y}$ and the target value of output $y$ i.e.,

$$\text{Loss, } \hat{L} = mse(y_{true} - \hat{y}) = \frac{1}{N_{batch}} \sqrt{\sum_{n=1}^{N_{batch}} (y_{true,n} - \hat{y}_n)^2} \qquad (11)$$

For the aeroelastic and sloshing problems, the loss function computes the squared differences between the network projected outputs and expected output in a *MIMO* (multiple-input multiple-outputs) respectively as follows.

$$\hat{L} = mse(y_{true} - \hat{y}) = \frac{1}{N_{batch}} \sqrt{\sum_{n=1}^{N_{batch}} \left[ (C_{L,n} - \hat{C}_{L,n})^2 + (C_{M,n} - \hat{C}_{M,n})^2 \right]} \qquad (12)$$

$$\hat{L} = mse(y_{true} - \hat{y}) = \frac{1}{N_{batch}} \sqrt{\sum_{n=1}^{N_{batch}} \left[ (F_{x,n} - \hat{F}_{x,n})^2 + (F_{y,n} - \hat{F}_{y,n})^2 + (M_{z,n} - \hat{M}_{z,n})^2 \right]} \qquad (13)$$

The model is trained for temporal sequences of inputs using *backpropagation* algorithm [25] by minimizing the losses using *Adam Optimization* algorithm [26]. The network is trained using data from the desired physical systems. Once the network weights and biases are established from the training, these parameters are stored and used subsequently to form a surrogate predictive model for various arbitrary future test data that are not present in the initial training data set. In this way, the computational cost for generating CFD data for training the network itself is a one-time process and need not be repeated often. The reliability of the prediction of arbitrary input data can be assessed by comparing with CFD computations. In this work the open source *Tensorflow* [27] libraries (version 1.10) are used to implement and train the RNN architecture. The training and testing data obtained from high-fidelity CFD simulations are imported and pre-processed for facilitating tensor manipulations in *Tensorflow* The hyperparameters of the RNN which include the number of epochs, training batch length and number of neurons in each hidden layer are optimized based on mean squared error of the predicted outputs for both networks.

The surrogate models for the prediction of aero-structural loads require motion information of the coupled system. The structural displacements for the forced motion of the airfoil is known beforehand and the aerodynamic and sloshing model surrogates predict the corresponding loads at each structural time step. When the airfoil is set free, the aerodynamic and sloshing load interaction with the structural motion is governed by Eqn. (7). The displacements computed from the solution of Eqn. (7) are fed back into the surrogate models as new inputs and the prediction marches on in time to yield the plunge and pitch displacements. This flowchart showing coupling of aero-structural model for both forced and free aeroelastic motion is shown in Fig. 7.

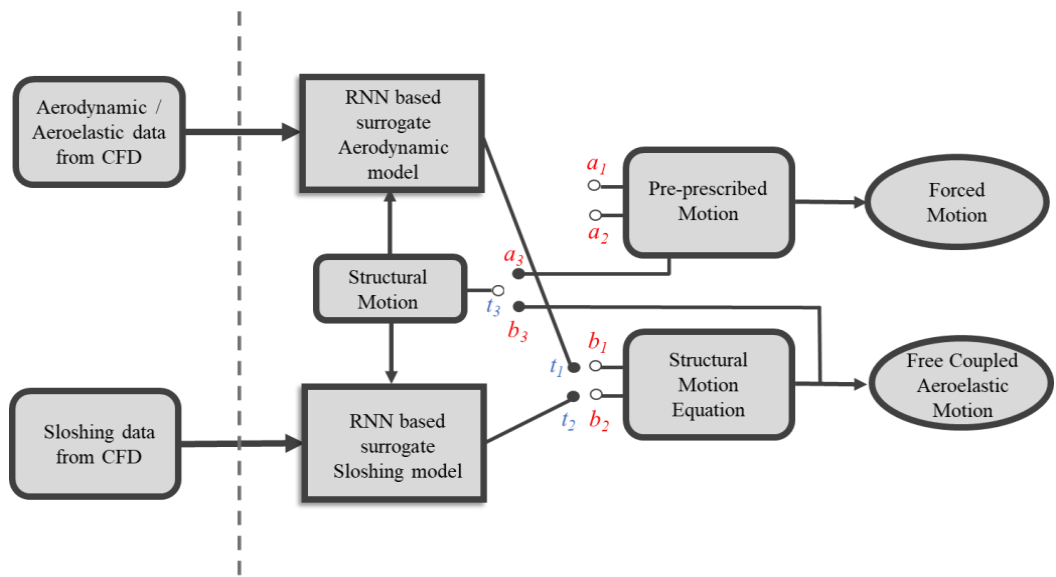

**Figure 7: Flowchart showing the coupling of aero-structural model for both forced and free aeroelastic motions**

The motion of aero-structural system must be selected to either forced predefined motion or free aeroelastic motion as shown in Fig 7 by connecting surrogate toggles $t_1$ and $t_2$ to toggles $a_1$ and $a_2$ or $b_1$ and $b_2$, respectively. The motion parameters, $h$ and $\alpha$, are supplied to the surrogates through $t_3$ via $a_3$ or $b_3$, depending on the choice of model. This determines the usage of either forced motion surrogates or free motion surrogates, as well as, the source of aero-structural motion parameters, i.e. $h$ and $\alpha$. For forced motion, $h$ and $\alpha$ are predetermined and the surrogate models essentially predicts the aerodynamic and sloshing loads respectively. For free motion, the structural motion parameters are computed by solving the aero-structural motion Eqns. (7)-(8) using the load predicted by the surrogate models. The structural motion sequences combined with loads predicted by the models are fed back in the surrogate model as inputs to produce subsequent predictions and to march the prediction in time. This machine learning based computational framework can be made more robust and richer for parametric variation with more training data over different flow conditions and motion amplitudes and frequencies.

## VI. Results and Discussions

An important aspect of the present study is the reduction of computational cost for investigating the multi-physics simulations for assessing the effects of fluid sloshing in a fuel tank embedded inside a wing section in a free pitch and plunge motion on the aeroelastic response of the system, while preserving the unsteady nonlinear features of both internal sloshing flow in the fuel tank and the external aerodynamic flow. The choice of training data and training methodology is a critical step for this problem as it must efficiently capture the dominant dynamics of the underlying physical system. Open source CFD code *SU2* is used for generating training data for aeroelastic prediction model, while *OpenFOAM* is used for generating training data for multiphase sloshing prediction model. The data generated from these solvers are validated against results in existing literature before training samples are computed and used for training. The wing section considered for this study corresponds to the NACA0012 airfoil immersed in an external inviscid transonic freestream flow at $M_\infty$ of 0.70 for which the flutter speed index, $V_f$ is 0.425. The computational domain consists of an O-type grid with farfield located at 25 chord lengths from the tip of the airfoil having about 11,000 mesh elements. The upper and lower airfoil surfaces are divided into 100 non-uniformly spaced grids symmetric about the chord. The fuel tank is embedded in the airfoil with its geometric centre coinciding with the elastic axis of the airfoil, 0.25 chord lengths from the leading edge, with tank length and height of about 0.30 and 0.09 chord lengths of the airfoil respectively. The computational domain for multiphase sloshing simulations consists of the enclosed fuel tank with its length and height divided uniformly into 150 and 125 points respectively resulting in 18,750 cells. Mesh convergence studies have been done on all the meshes used in the present study and only converged solutions are used as the training data as well as the ground truth data.

**I. Validation of *SU2* Aero-Structural Solver:**

The high-fidelity solutions computed using *SU2* are validated for steady and unsteady computations. Steady inviscid transonic flow field around a NACA0012 airfoil immersed in a free stream Mach number $M_\infty$ of 0.80 and at an angle of attack, $\alpha$ of 1.25° is computed. The Mach contours and the coefficient of pressure, $c_P$, distribution on the lower and upper airfoil surfaces are shown in Fig 8 (a) and Fig 8 (b) respectively. For unsteady flow validation, the airfoil is immersed in a flow field with freestream Mach number $M_\infty$ of 0.60 and then from the steady state flow it is forced to pitch according to the following relation,

$$\alpha = \alpha_{mean} + \alpha_0 \sin(\omega t) \quad (14)$$

where, $\alpha_{mean}$ is 2.89°, $\alpha_0$ is 2.41° and $\omega$ is 32.9952 (rad/s). The variation of the computed lift and moment coefficients with pitching angle are shown in Fig 8(c) and Fig 8(d) respectively. The aerodynamic solver for steady flow in the present study is validated against the JST [28] solution scheme in *SU2*, and the unsteady solution is validated against that of MacMullen et al. [29]. As the computed results show good agreement with the published results, the reliability of the flow solver in generating steady and unsteady inviscid external transonic flow field data for training samples and for unsteady aeroelastic computations is thus validated.

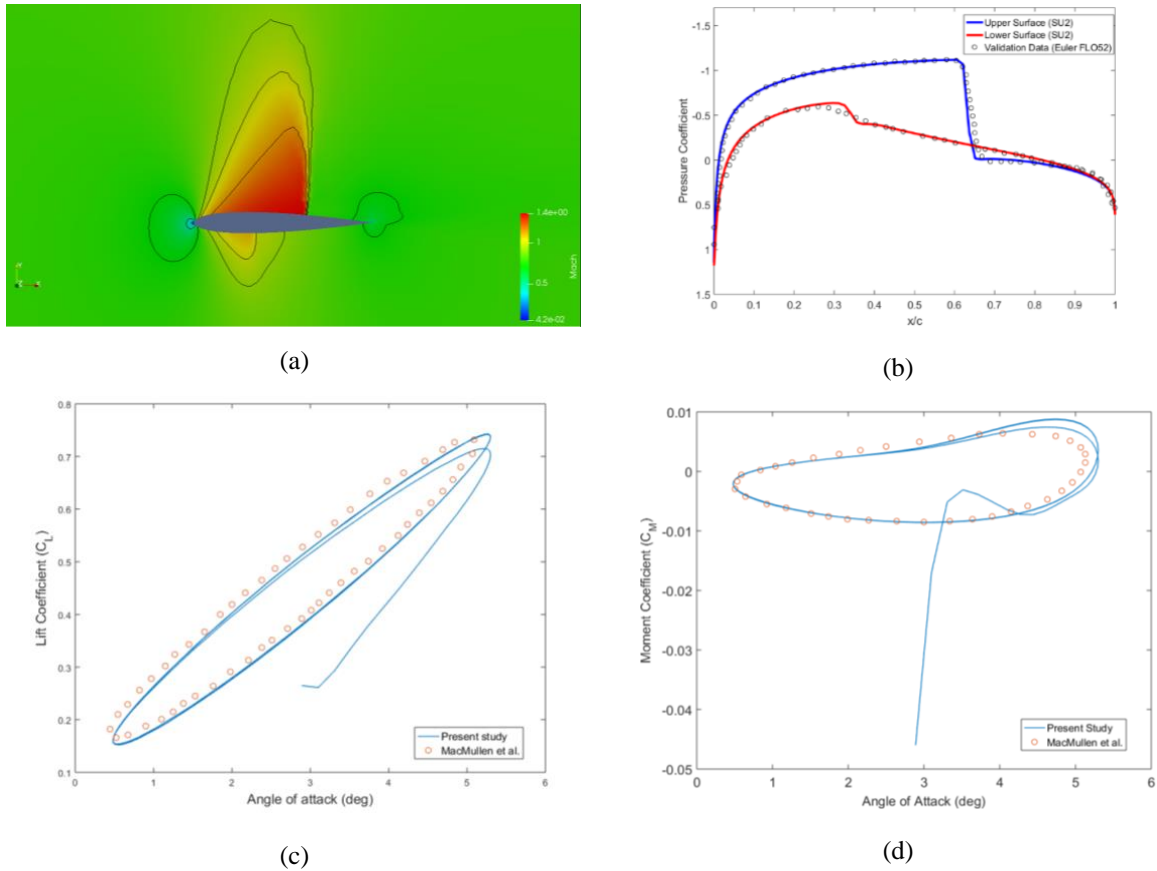

Figure 8 (a) Mach contours of NACA0012 airfoil at M$_\infty$=0.80 and α = 1.25°, (b) Coefficient of pressure distribution of the upper and lower airfoil surfaces, (c) Lift coefficient variation with angle of attack for harmonic pitching comparison of present study with published results, and (d) Moment coefficient variation comparison with published data for the same motion

**II. Validation of *OpenFOAM* Solver:**

The *interDyMFoam* library within *OpenFOAM* is used for computing the two-phase fluid sloshing in a rectangular tank with internal dimensions of lateral length (*L*), height (*H*) and breadth (*B*) of 1000 *mm* by 980 *mm* by 100 *mm* respectively with a 40% fill-level. The computational domain for the validation is the moving tank, having 1000 uniformly divided mesh cells in the lateral direction, 400 cells in the vertical direction and 100 cells in the normal direction. The tank is set to a transverse motion about *x*-axis with a frequency of 1.0059 *Hz* and an amplitude of 0.0145 *m*. The dynamic pressure and forces computed from the VOF model are compared with experimental and numerical data of Yusong et al. [30] in Fig 9.

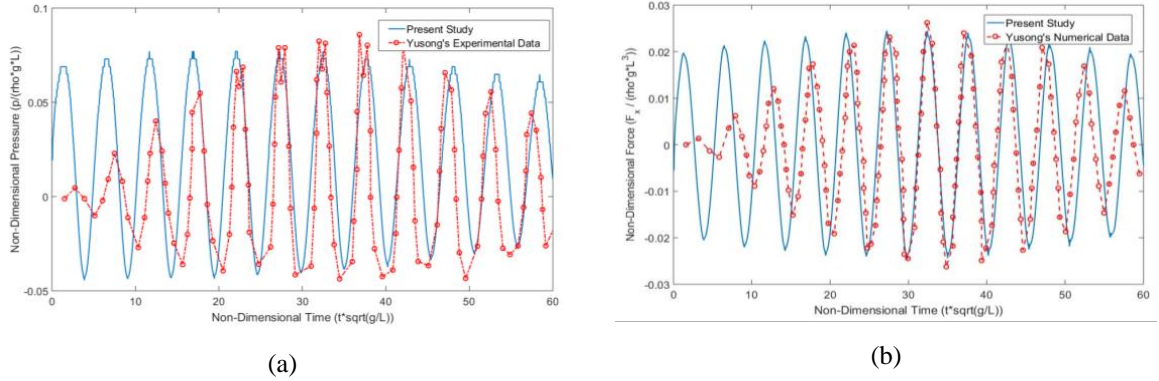

(a)                 (b)

**Figure 9.** Comparison of (a) dynamic pressure and (b) transverse sloshing forces obtained from present study with experimental and numerical measurements respectively of Yusong et al. [28]

It can be seen that the computed dynamic pressure and sloshing forces estimated from the current numerical model follow the trends shown in the experimental data for the traverse motion of the partially filled tank in the latter part of the time series. The mismatch in the initial pressure and force response can be attributed to presence of transient noise in the data. Furthermore, it must be noted that the data from literature has been manually digitized from the published figures and hence may have some measurement errors at certain points. Within these limits the validation of the computed sloshing loads can be concluded as being acceptable.

### III. Single Step Prediction Test – Aero-Structural Surrogate Model

The aero-structural surrogate model is required to predict $C_L$ and $C_M$ at the current time step given the information of the $h$ and $\alpha$, as well as the aerodynamic loads and structural displacements information from previous time steps. The unsteady airfoil motion is initiated with a forced pitching of the airfoil immersed in an external steady transonic flow field and after a few cycles of periodic variation of the aerodynamic loads, the airfoil is set to free unsteady motion which is driven by the interaction between the free structural motion and the ambient external flow field. To encompass all the possible motions of the airfoil, the prediction network is tested for both unsteady forced and free airfoil motion.

**(a) Forced Pitching of the Airfoil:** The airfoil is immersed in a uniform free stream flow at $M_\infty$ of 0.70 and is forced to pitch according to $\alpha = \alpha_{mean} + \alpha_0 \sin(\omega t)$ with an angular frequency $\omega$ of $2\pi$ (rad/s) about the angle of attack $\alpha_{mean}$ which is assumed 0° and the pitching amplitude $\alpha_0$ is 2°. Corresponding to the forced pitching shown in Fig. 10 (a), the computed time variation of lift ($C_L$) and moment ($C_M$) coefficients and from CFD simulation are shown in Fig. 10 (b). The network is trained using randomly selected input data consisting of a matrix of pitch angle, plunging displacement (all zeros in this case) and lift and moment coefficient variation with time with a batch length of 40 for each epoch. An example of one such training instance is shown in Fig. 10 (a) - 10 (b) with black dots (i.e. corresponding to the range of 50 to 89 time steps in the training data). The network consists of two hidden layers consisting of 120 and 80 neurons, respectively. The training errors are estimated using the mean squared error (*mse*) of predicted and expected values for the batch sequences of outputs, in this case lift and moment coefficients. The network weights and biases are learnt during the network training in order to minimize the loss function, i.e. Eqn (12), the convergence of which is shown in Fig 10 (c). For the training instance considered the convergence of the *mse* is of the order $O(10^{-5})$, which is achieved after 1500 epochs as shown in Fig 10 (d). Once trained, the network can be tested to assess its predictive reliability for one time step ahead for any sequence of inputs.

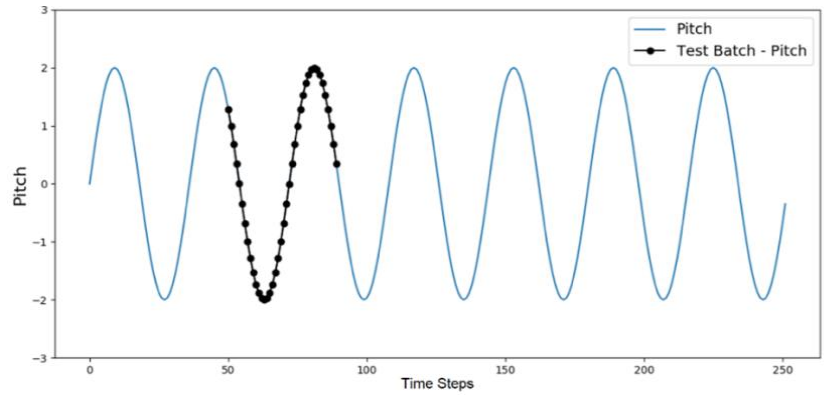

(a)

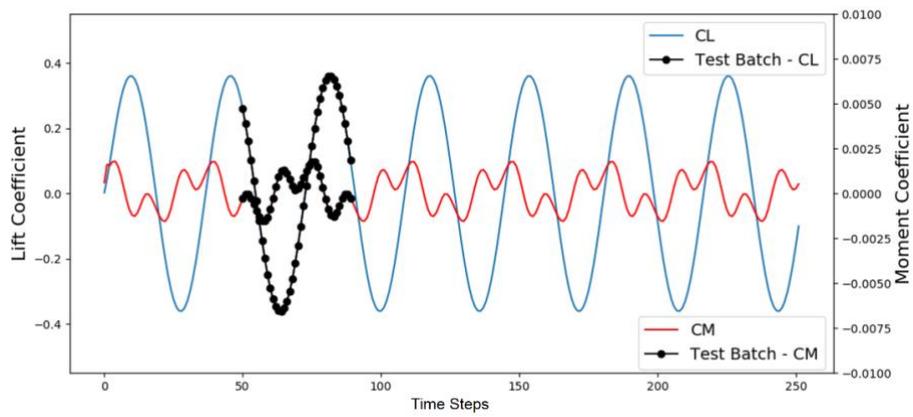

(b)

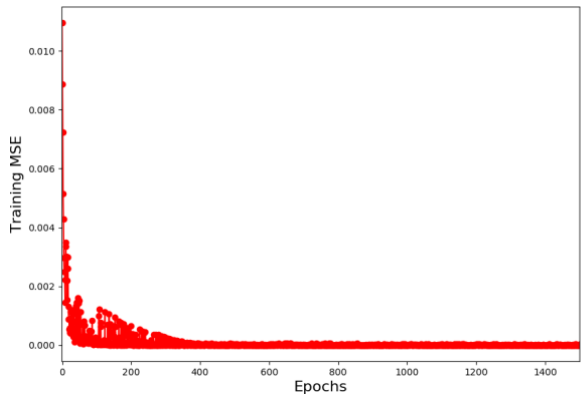

(c)

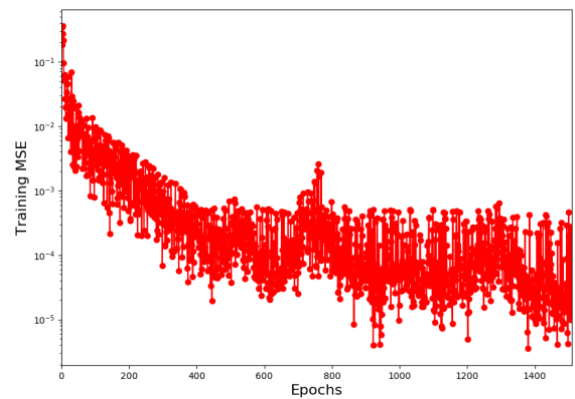

(d)

**Figure 10: (a) Pitching displacements of the airfoil in free motion with a random batch of inputs for testing the RNN, (b) Corresponding variation of lift and moment coefficients; Variation of (c) MSE and (d) MSE (in log scale) vs. epochs during network training for aerodynamic data.**

Once trained, the RNN is tested with a batch of input sequence for demonstration of prediction accuracy. The input sequences shown in Fig. 10 (a) – 10 (b) with black dots, corresponding to the range of 50 to 89 time steps in the training data is selected. These input sequences are fed into the RNN to predict the lift and moment coefficient at the next time step, i.e., the $90_{th}$ time step in the training data. Figures 11 (a) and 11 (b) show an enlarged view of the training instances (labelled as grey points) corresponding to the batch length shown as black dots in Figs. 10 (a) and 10 (b) (i.e. input sequence), expected output (i.e. target) and actual network outputs (i.e. predicted) respectively for both aerodynamic lift and moment coefficients. The target prediction at the $90_{th}$ time step is shown with a blue dot and the actual network prediction is shown with a red dot.

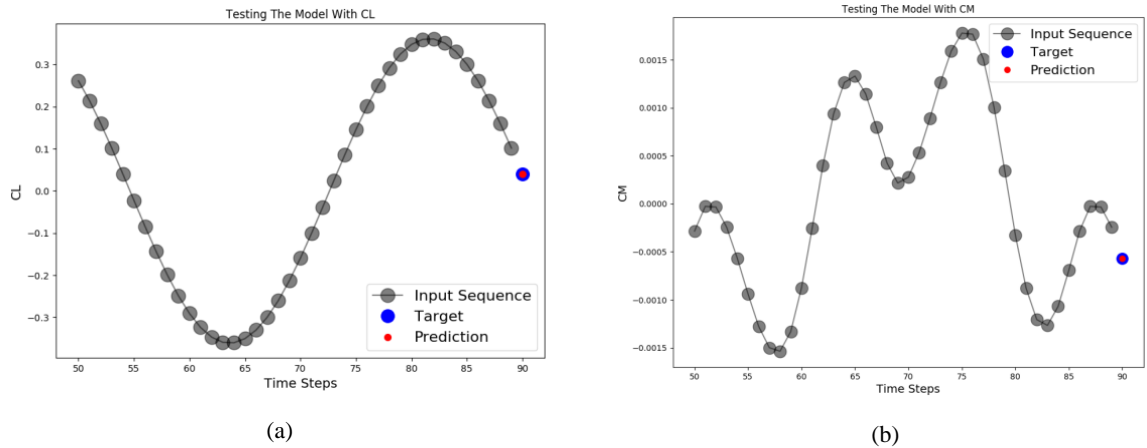

(a)                                             (b)

**Figure 11: The training instance, the expected output and the network prediction for lift coefficient using randomly chosen batch for (a) lift and (b) moment coefficients**

The lift and moment coefficients at one time step ahead of the input series (give value of *t*) predicted by the RNN (labelled as red dot in Figs. 11 (a) – 11 (b)) is compared with the target values computed from CFD simulation (labelled as blue dot in Figs. 11 (a) - 11 (b)). The defined acceptable error for single step prediction is of the order $O(10^{-5})$. The accuracy of prediction is measured by computing squared errors for $C_L$ and $C_M$ at time step of interest, which is achieved. This is expected as the RNN has been trained to reduce *mse* to this order, shown in Fig 10 (d). The prediction accuracy of the RNN can be improved arbitrarily further by tuning the hyperparameters.

**(b) Free Plunging and Pitching of NACA0012:** After a few cycles of forced pitching and plunging of the airfoil initialized with the flow field at free stream $M_\infty$ of 0.70 at 0° angle of attack, the airfoil is set free to move so that the pitch and plunge displacements are determined via the interaction with the structural dynamics and ambient unsteady aerodynamic flow. Fig. 12 (a) shows the time history of plunging and pitching displacements of the airfoil and Fig. 12 (b) shows the corresponding lift and moment coefficients. Fig 12 (c) shows the convergence of weights and biases for the present training instance in form of reduction in *mse*, which converges to the order $O(10^{-5})$ after 1000 epochs, as shown in Fig 12 (d). This airfoil displacement and aerodynamic loads data are used as the training set for one step aeroelastic motion prediction model.

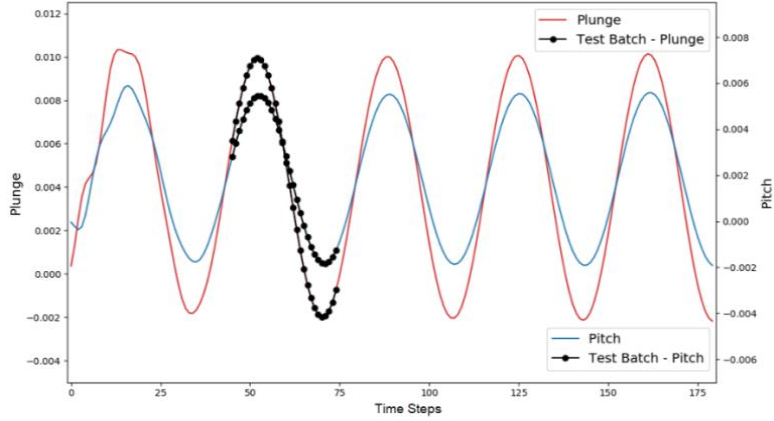

(a)

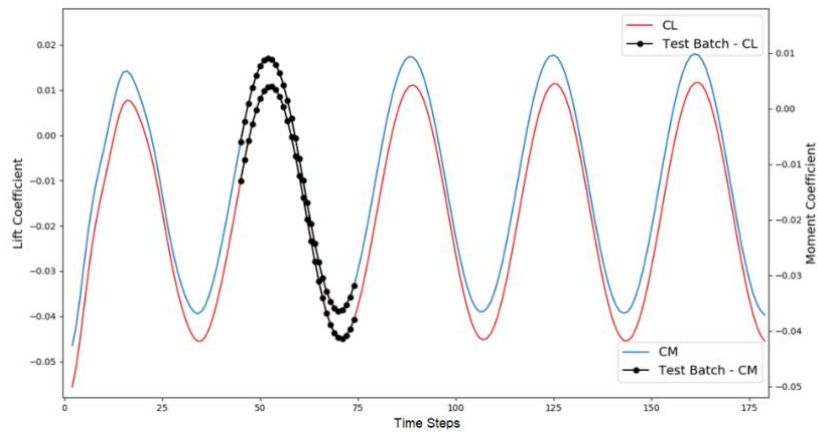

(b)

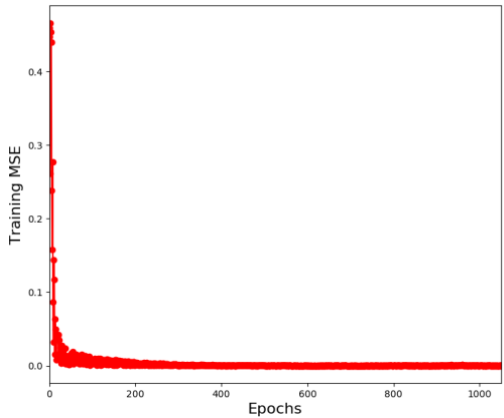

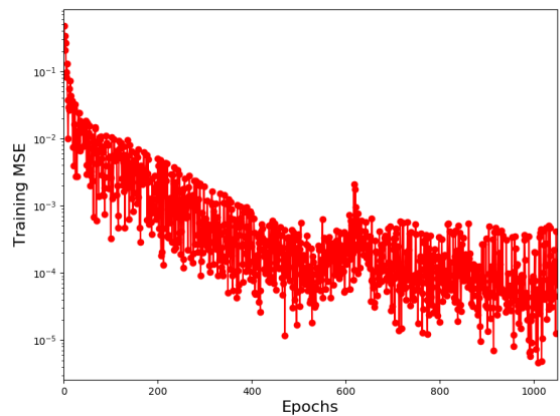

(c)                                                                 (d)

**Figure 12: Time history of (a) plunging and pitching displacements of the airfoil in free motion with a random batch of inputs for testing the RNN, (b) Corresponding lift and moment coefficients, Variation of (c) MSE and (d) MSE (in log scale) vs. epochs during network training for aerodynamic data.**

The RNN architecture is trained on free plunging and pitching data and the learnt weights and biases are used for testing its prediction accuracy. The training inputs are sequences of data from airfoil motion series in the form of plunging and pitching and the corresponding aerodynamic loads $C_L$ and $C_M$. The input sequences are selected randomly for each training epoch; an instance corresponding time steps ranging from 45 to 74 is labelled with black dots in Fig 12 (a) and Fig 12 (b). These inputs sequences are fed into the RNN to check the prediction accuracy for one time-step in future. Enlarged views of the training instance, expected output and actual network outputs are shown in Fig 11 (a) and Fig 11 (b), respectively. The trained network is tested with a batch of input sequences for lift and moment coefficients shown as grey dots in Fig. 13 (a) - 13 (b) corresponding to time steps ranging from 45 to 74 of the training data.

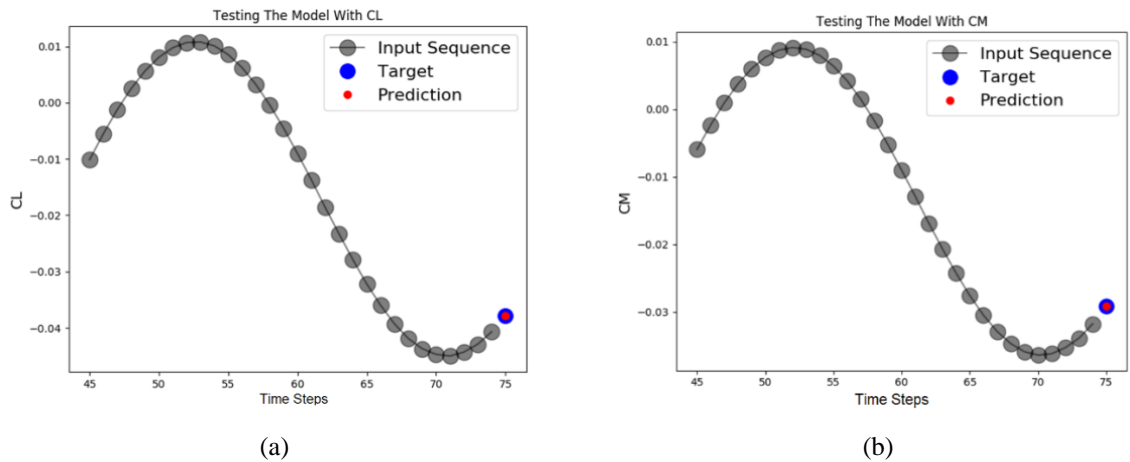

(a) (b)

**Figure 13: (a) The training instance, the expected output and the network prediction for lift coefficient using randomly chosen batch, and (b) Same parameters shown for moment coefficient**

The target prediction at the 75th time step is shown with a blue dot, and the actual network prediction is shown with a red dot. The prediction accuracy is estimated by computing the squared error of the outputs at the time step of interest. The error is found to be of the order $O(10^{-4})$ for the aforementioned testing, which is expected as the *mse* is reduced to order between $O(10^{-4})$ and $O(10^{-5})$ during training, as shown in Fig 12 (d).

**IV. Single Step Prediction Test for RNN Surrogate Model for Tank Fuel Sloshing Loads Prediction**

The RNN surrogate prediction model for fuel sloshing in the tank aims to predict *lateral* and *vertical* forces on the tank walls arising from fuel sloshing and the *moment* generated by these forces about the geometric center of the tank at the present time step. As the tank undergoes a forced pitching and plunging, the input for the model consist of the information of plunge, *h* and pitch, *α*, at the previous and current time step and also the information of $F_X$, $F_Y$ and $M_Z$ at the previous time step. The data used for training and testing RNN is scaled to values between 0 and 1, as input data sequences in same order facilitates efficient training of weights and biases. Figure 14 (a) shows the normalized temporal history of the pitch and plunge motion and Fig. 14 (b) shows the corresponding sloshing forces and moments.

The network is trained where inputs are the temporal sequences of tank motion in form of plunging (*h*) and pitching (α) at the present and previous steps, and time history of previous sloshing loads $F_X$, $F_Y$ and $M_Z$, respectively. The network outputs are $F_X$, $F_Y$ and $M_Z$ at the present time step. The network architecture contains two hidden layers, containing 170 and 120 neurons each. The total number of neurons used for this case are more than that used for the pitching and plunging airfoil aerodynamics in Section VI.II(a) because there are a greater number of inputs and outputs and the outputs have more features than those of pitching and plunging airfoil aerodynamics. A batch length of 15 previous time step data is used for the inputs and *loss* is computed as per Eqn 13 for subsequent epochs during the training process. Fig 14 (c) and Fig 14 (d) shows the convergence of weights and biases for the present training instance in form of reduction in *mse*, which converges to the order $O(10_{-5})$ after 4000 epochs, as apparent form Fig 14 (c) and Fig 14 (d).

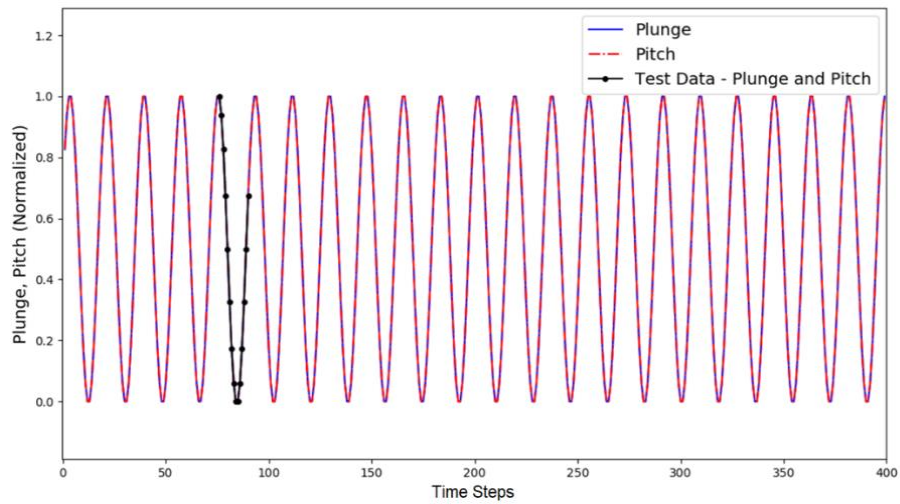

(a)

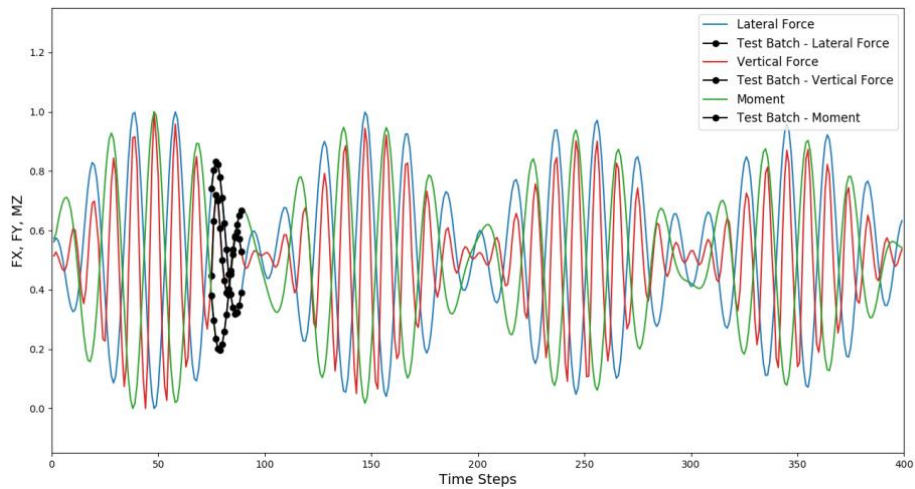

(b)

**Figure 14: (a) Pitch and plunge motion time series structural inputs for tank motion, (b) Sloshing loads in response to tank motion,**

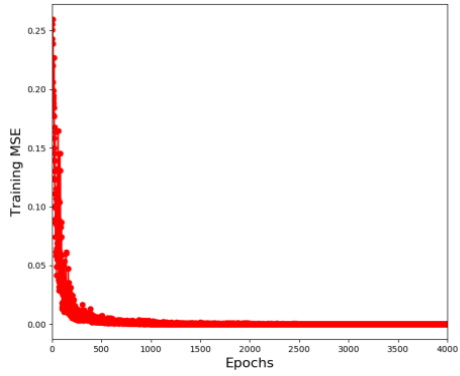
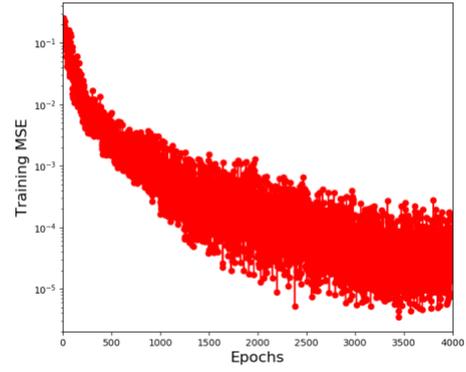

(c)                                                                (d)

**Figure 14: (Continued) Variation of (c) MSE (d) MSE (in log scale) vs. epochs during network training for aerodynamic data.**

To evaluate and demonstrate the prediction accuracy of the trained RNN, input sequences of plunging and pitching motions shown with black dots in Fig 14 (a), and corresponding $F_X$, $F_Y$ and $M_Z$, shown in Fig 14 (b) are selected from the training data and is fed into the RNN. The input sequences ranges from 75 to 89 time steps of the training data to evaluate the sloshing loads prediction at 90th time step. Zoomed-in comparison of training data, expected values and predicted values model outputs namely, *lateral forces*, *vertical forces* and *moment* respectively represented by grey dots, blue dot and red dot, respectively, are shown in Fig 15 (a) - (c)

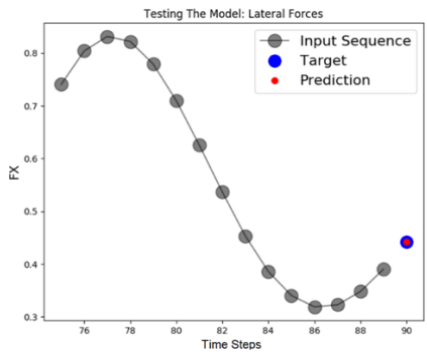
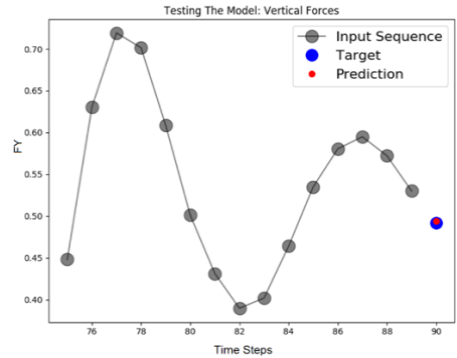

(a)                                                                (b)

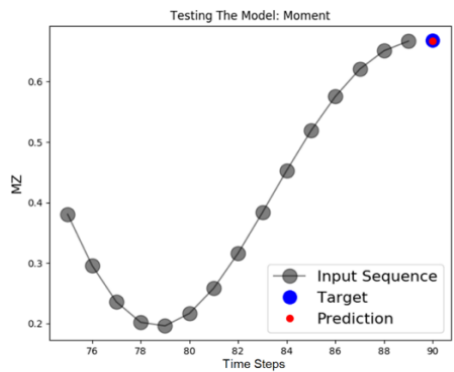

(c)

**Figure 15: The training sample, expected values and prediction values of surrogate model for, (a) lateral force, (b) vertical force, and (c) moment**

The prediction error is computed by taking squared difference of the predicted value and the CFD data for $F_X$, $F_Y$ and $M_Z$ at the time step of interest. The squared error is found to be of the order $O(10^{-4})$. This is expected as the RNN has been trained to reduce the *mse* between the order $O(10^{-4})$ and $O(10^{-5})$, as apparent from the *mse* shown in Fig 14 (d). This range of prediction error is deemed acceptable for the present study. However, the prediction accuracy of the RNN can be improved further by tuning the hyperparameters of the network.

**V. Prediction of Aeroelastic Responses Using RNN**

On the basis of the accuracy of RNN surrogate prediction of aerodynamic load coefficients and sloshing loads at one time step ahead of the input sequences, in this section the surrogate model is used to predict the system responses for a longer time by considering a case of a diverging aeroelastic response of the wing section. This is facilitated by appending the network output obtained at the current time step to the sequence of its previous values and then using it as inputs for prediction at the next time step as outlined in Fig.7.

**(a) Aeroelastic Response Prediction:** For this case, a higher freestream Mach number is chosen for free motion, which in turn has a more diverging aeroelastic response for testing the robustness of the RNN prediction. The surrogate prediction model is built on the RNN architecture for the aeroelastic response of a NACA0012 airfoil which is forced to pitch and plunge for two cycles from a freestream flow at $M_\infty$ of 0.80 and then set free to pitch and plunge. The flutter speed index $V_f$ for this case is 0.425. The pitching and plunging motion, lift and moment coefficients and the corresponding phase plots for pitching and plunging are computed on the basis of free pitching and plunging motion of the airfoil are shown in Figs 16 (a) - (d),

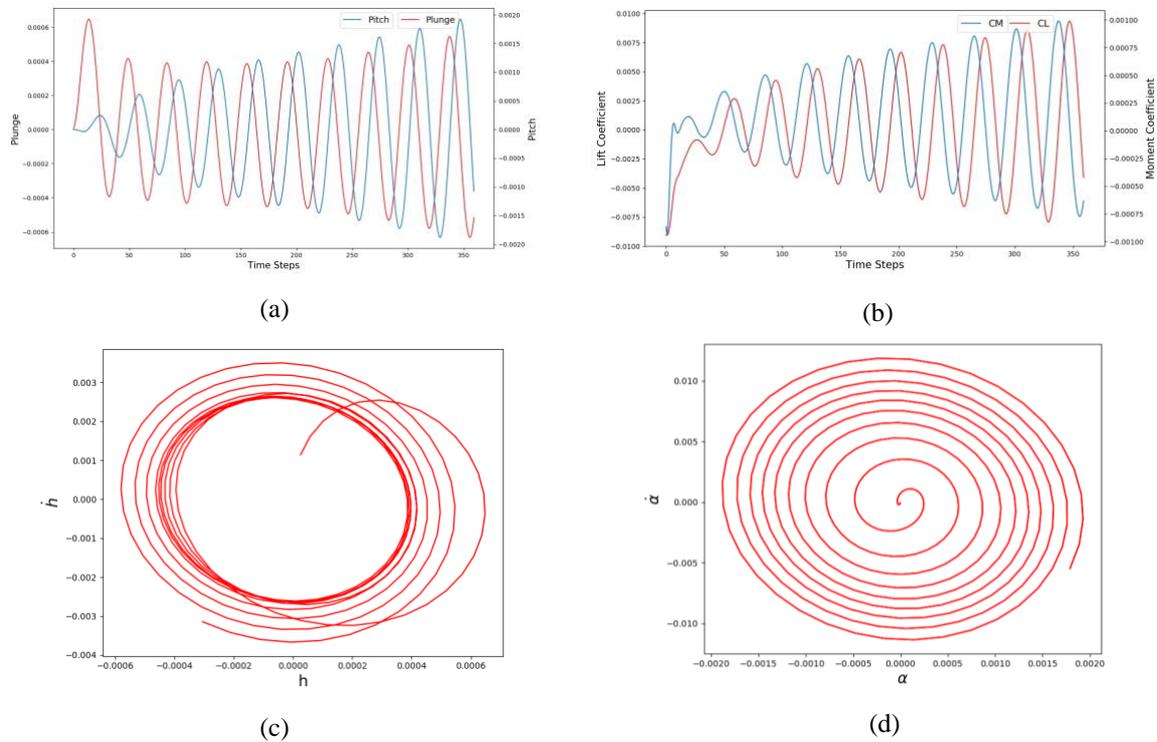

**Figure 16: (a) Plunging and pitching displacements of the airfoil in free motion, (b) corresponding lift and moment coefficients obtained for the free motion of the airfoil (c) phase plot for plunging, and (d) phase plot for pitching**

The computed high-fidelity CFD dataset is first divided into training (red color) and testing (blue color) data sets as shown in Fig 17 (a) and Fig 17 (b) which show respectively the plunging and pitching data sets. A train-test data ratio of 1:1 is maintained for all other inputs as well, namely, $C_L$ and $C_M$. The RNNs are trained using the data shown in red for a combined plunging and pitching motions and their corresponding lift and moment coefficients. Once trained, the surrogate model will be tested for prediction accuracy using the testing dataset shown in blue.

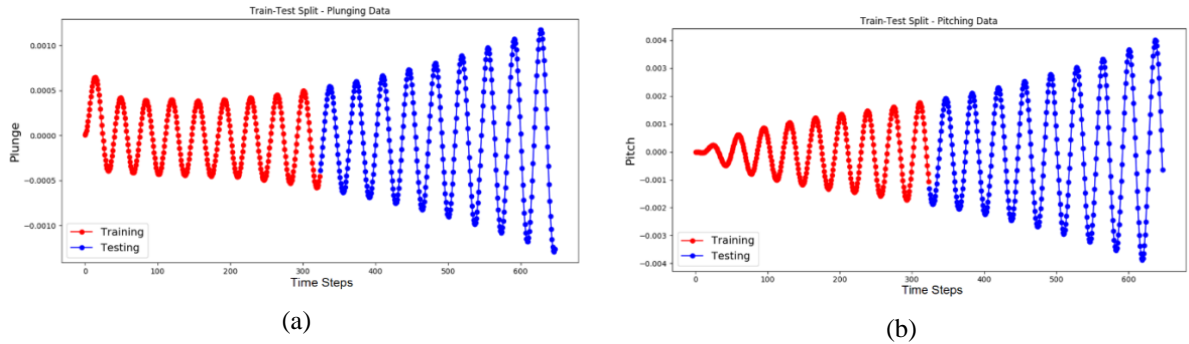

(a)  (b)

**Figure 17: (a) Plunging data divided into test-train sets for free-aeroelastic motion of NACA0012 airfoil, and (b) pitching data divided into test-train sets for the same motion**

The RNN architecture with four input sequences feeding through two hidden layers of neurons using *ReLU* activation function generating two outputs as shown in Fig 6 (a) is trained with input sequences consisting of plunge and pitch displacement data and the corresponding lift coefficient ($C_L$) and moment coefficient ($C_M$) data with batch length of 30. These training data for lift and moment coefficients are shown in red and green lines respectively in Fig. 18 (a). The predicted variation of lift and moment coefficients from the trained model when temporal sequences of displacement parameters $h$ and $\alpha$ and $C_L$ and $C_M$ from the testing data set are fed into the network as shown as blue and purple color respectively in Fig 18 (a). The predicted $C_L$ and $C_M$ from the RNN, represented by yellow dots, are overlaid on the ground truth CFD data for comparison. The predicted variation of the aerodynamic lift and moment coefficients vs. plunging and pitching amplitudes are compared against CFD data in Figs 18 (b) – (e).

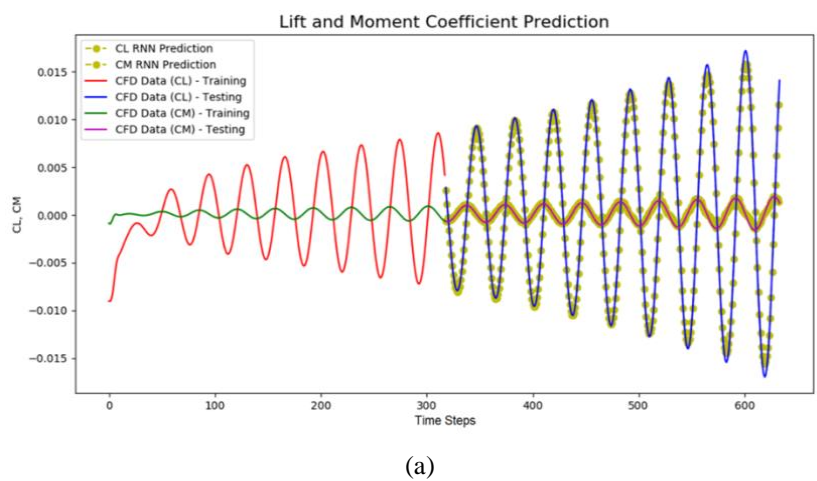

(a)

**Figure 18: (a) Airfoil displacements and lift and moment coefficients as training data inputs, testing targets and surrogate model prediction outputs,**

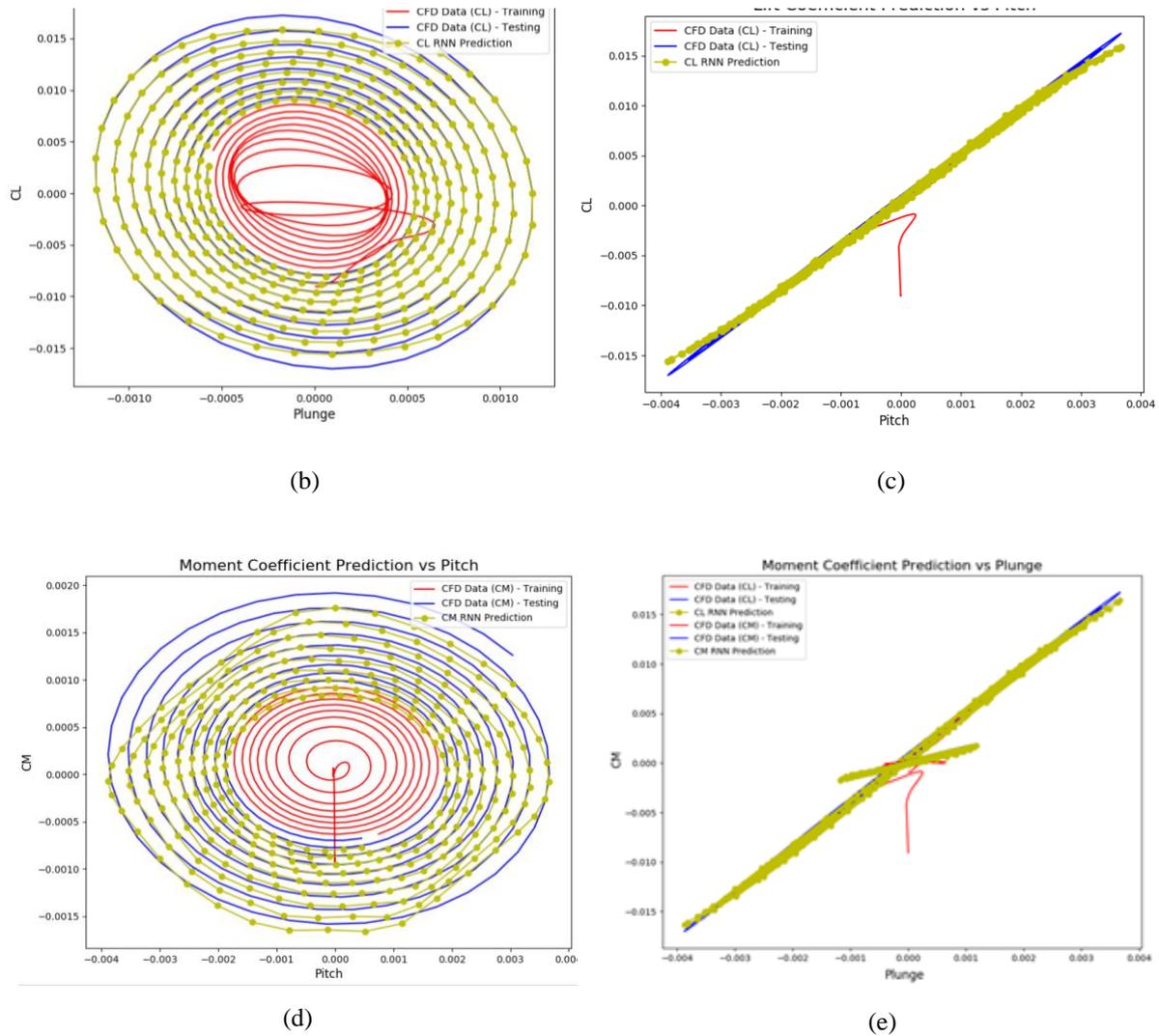

**Figure 18:** (Continued) (b) phase plot of comparison of lift coefficient against plunge, (c) phase plot comparison of moment coefficient against pitch, (d) phase plot comparison of moment coefficient against pitch, and (e) phase plot comparison of moment coefficient against plunge

The surrogate model for aeroelastic load prediction based on motion parameters, $h$ and $\alpha$, predicts the lift and moment coefficients for testing data set to an acceptable accuracy. The model slightly underpredicts the lift coefficient after the airfoil goes into violent pitching motion around the $500_{th}$ time step as apparent from Fig 18 (a). This is because the model is trained only using first half of the aeroelastic motion time series and it had not encountered higher values of pitching and plunging data to accurately predict corresponding lift and moment coefficients. The variation of $C_L$ and $C_M$ with $h$ and $\alpha$ is shown in Fig 18 (b) – 18 (e) which clearly shows the underprediction of aerodynamic load coefficients as the aeroelastic motion diverges. Such inaccuracies in prediction can be eliminated by using a richer training dataset containing dynamics of aeroelastic motion or finely tuning the hyperparameters of the RNN.

**(b) Sloshing Loads Prediction:** The partially filled fuel tank moves in synchronization with the airfoil and hence the sloshing loads are computed using structural inputs provided by the aeroelastic equations. The current state of fluid in tank also depends on the previous states, which are represented by temporal sequences of loads due to fluid sloshing on the tank walls. The complete time series of the sloshing loads can be generated by appending one-step predictions from the predictive model over the length of complete time series.

The tank is excited with structural motion obtained from free motion of NACA0012 airfoil experienced after forcing the airfoil for 2 cycles of pitching motion and then letting it move freely with freestream $M_\infty$ of 0.60 and speed index $V_f$ of 0.425. The free motion of the airfoil is same as in full time series generation of aeroelastic case, shown in Fig. 17 (a) and Fig. 17 (b). The time series is divided in training and testing data with 1:1 ratio, latter part being utilized for testing. The RNN architecture is remains same as chosen in Section C. III. The corresponding horizontal and vertical forces on tank walls, and the corresponding moments are the outputs of the predictive model, are shown in Fig. 19.

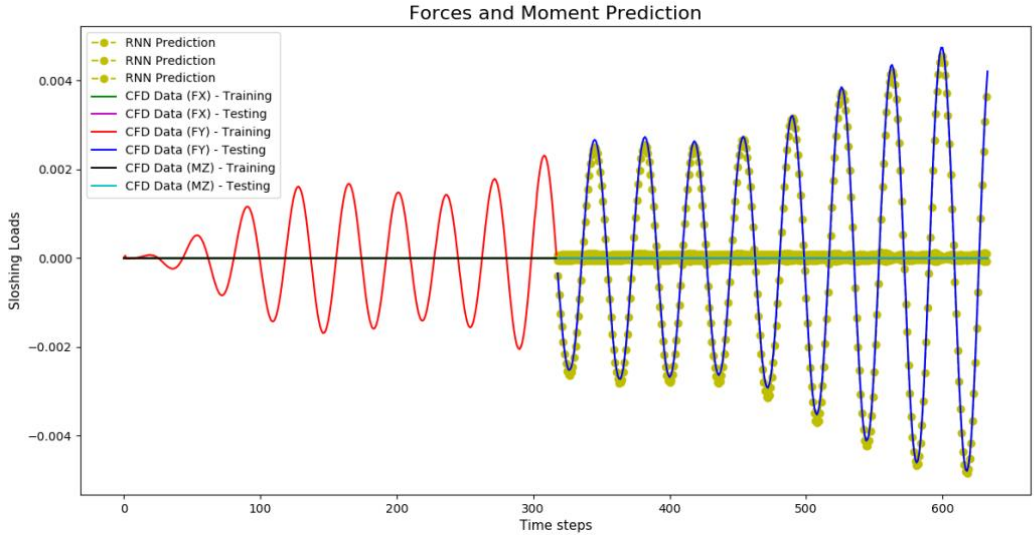

**Figure 19: Sloshing loads due to tank motion in form of forces and moments as training data inputs, testing targets and surrogate model prediction outputs**

The surrogate model for sloshing loads prediction based on $h$ and $\alpha$, predicts the vertical and horizontal forces and moment for testing data set to an acceptable accuracy. The RNN predictions represented by yellow dots in Fig 17 are superimposed on the CFD data used for testing for all outputs, namely $F_X$, $F_Y$ and $M_Z$. The model accuracy can be improved arbitrarily by tuning the hyperparameters of the RNN.

**(c) Coupled Aeroelastic Motion with Sloshing Tank:** The aerodynamic and sloshing loads on the wing section are predicted by the surrogate model and the resulting aeroelastic motion is computed using Eqn 7. The computed structural motion is fed back into the surrogate models to generate loads for the next time step, as elaborated in the flowchart in Section V. The effect of sloshing on effective lift and moment coefficient of the coupled aero-structural system is depicted in Fig. 20 (a) and (b) respectively. The structural motion of the airfoil as computed using predictive surrogate models is compared with the structural motion computed using CFD in Fig 20 (c) and (d).

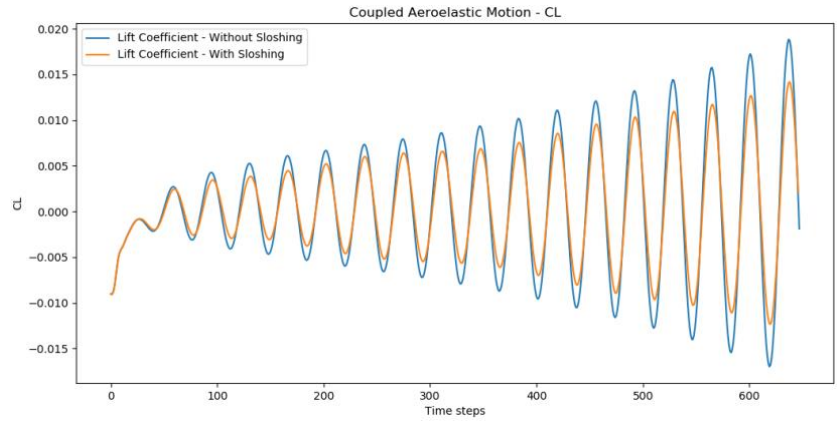

(a)

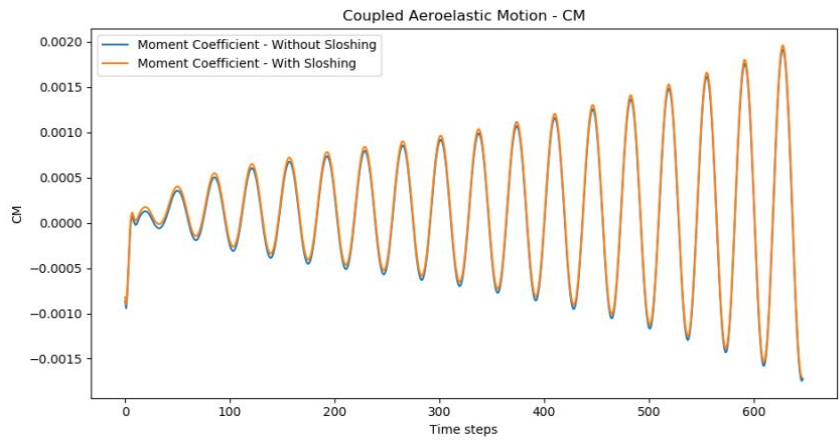

(b)

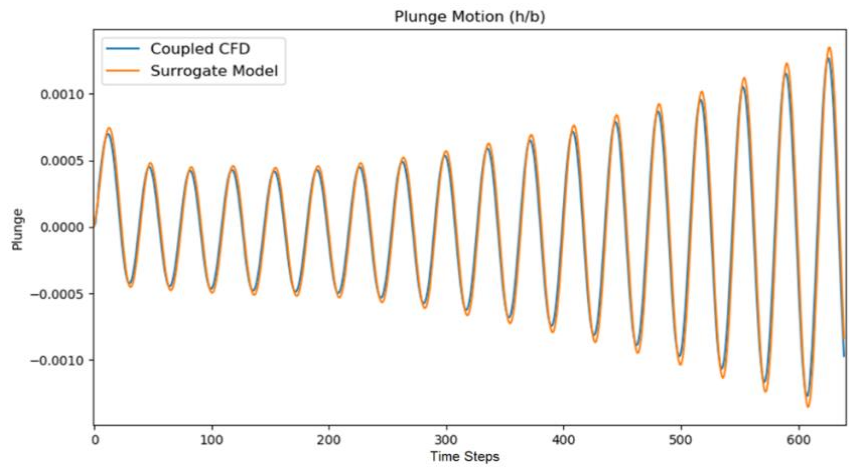

(c)

**Figure 20: (a) The effective lift obtained by coupling sloshing tank with the aeroelastic wing section using *RNN* based predictive model, (b) Corresponding moment coefficient, (c) Comparison of plunge motion of airfoil computed from surrogate model and high-fidelity CFD,**

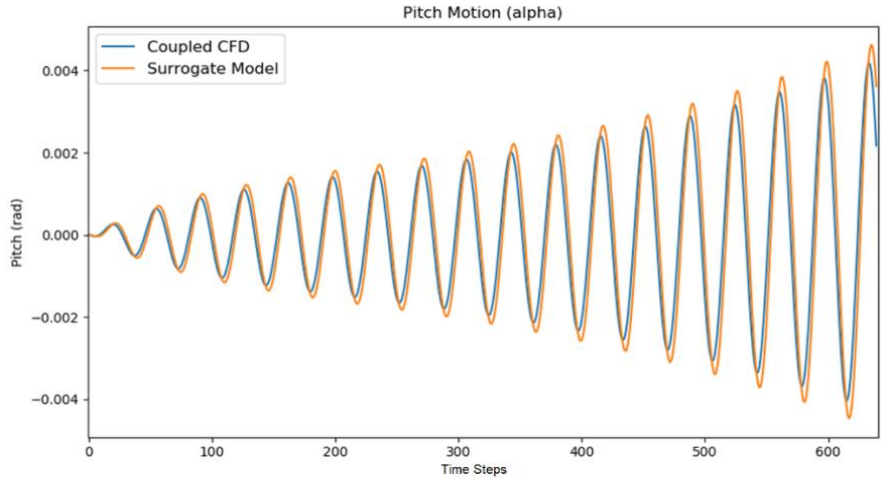

(d)

**Figure 20: (Continued) (d) Comparison of pitching motion of airfoil computed from surrogate model and high-fidelity CFD**

There is a significantly more effect of sloshing on the combined vertical loads due to aerodynamics and sloshing than normal moment. This is attributed to the fact that the centre of gravity of sloshing fluid does not move enough to cause effective moment on the elastic axis of the wing section. The structural displacements computed from the surrogate model and CFD are good agreement for initial free motion of the airfoil. The surrogate model slightly over-predicts the plunging and pitching motion when compared with CFD data. This mismatch can be attributed to the fact that training data for surrogate models does not contain larger airfoil displacements and hence lower prediction accuracy in that region.

**(d) Computational Cost Analysis of High-Fidelity and Surrogate Models:** The computational cost is measured in terms of CPU time of simulation running on an *Intel® Xeon® CPU E5-1650 V3 @ 3.50 GHz* processor running on single core and is summarized in Table 1.

Table 1. Computational Cost Analysis of High Fidelity CFD and Surrogate Predictive Model

|  | Time (in seconds) | Total Time (in seconds) |
|---|---|---|
| **CFD Simulations** |  |  |
| Aeroelastic | 558.010791 (forced) + 1875.305696 (free) | 3895.756308 |
| Sloshing | 1462.439821 |  |
| **Surrogate Predictive Model** |  |  |
| Aeroelastic | 67.407425 | 147.466771 |
| Sloshing | 80.059346 |  |
| **Computational Savings** | (3895.756308-147.466771) / 3895.756308*100 = **96.214682%** |  |

The RNN-based surrogate models for predicting aero-structural motion and sloshing loads in a partially filled fuel tank rigidly attached to the moving airfoil require one-time training using data obtained from CFD simulations. The computational time required to generate training data set is not considered for calculation of computational savings. The surrogate model developed in this work can be trained using rich data set containing multiple flow conditions and can be used to predict aero-structural loads for a wider variety of motions, flow conditions, and different fill-levels of fuel tank, thus further enhancing the utility and computational efficiency of surrogate model over high-fidelity CFD simulations.

In the light of recent development in physics informed neural networks as demonstrated by Raissi [31], the present study will be extended to explore the computational benefits of replacing solution of motion parameters from aeroelastic structural equation with a physics-augmented neural network. The complete machine learning computational framework is aimed to address quick prediction of coupled aerostructural motion to assess the effects of fuel sloshing on free aeroelastic motion.

## VII. Conclusions

This study has developed recurrent neural network (RNN) surrogate models to efficiently and accurately predict the unsteady aerodynamic loads of a free pitching and plunging airfoil in transonic flow field and also to predict the fuel sloshing loads in a partially filled rectangular tank subjected to the same displacements of the airfoil. The study also demonstrates the feasibility of coupling these two RNN surrogate models to predict the aeroelastic response of a two-DOF airfoil in which a partially filled rectangular fuel tank is embedded. These RNN networks are trained once with the high-fidelity CFD training sample data and the trained network parameters are stored for future reuse, thus eliminating the need of training. The training of the surrogates for external airfoil aerodynamics are computationally marginally cheaper than the surrogates for fuel tank sloshing loads owing to the increased number of outputs The coupled RNN surrogates are tuned and used to predict both forced and free motion of the coupled aero-structural-fuel tank system for routine prediction of subsequent motion at a much cheaper computational expense with minimal compromise of prediction accuracy. The training methodology and prediction accuracy of the surrogate models are tested with forced motion of the coupled aero-structural system where training and testing data has same amplitude of motion. The models are then trained on limited data set of free diverging aero-structural motion and tested on data set with more intense divergence. The unsteady aerodynamic and sloshing loads predicted by the surrogate models, coupled with structural motion equation is utilized to predict the free motion of the aero-structural system. The resulting motion and corresponding loads agree well with the CFD data, with a significant reduction of 96.21% in the computational cost. Such RNN surrogates show enormous potential for routine use for predicting the onset of flutter, limit cycle oscillations and for flutter mitigation directed aerodynamic wing design.